\documentclass[aps,prl,superscriptaddress,showpacs,twocolumn,longbibliography]{revtex4-2}
\usepackage[left=1.9cm,right=1.9cm,bottom=1.9cm,top=1.9cm,ignoreall]{geometry}
\usepackage[utf8]{inputenc}
\usepackage{graphicx}
\usepackage{amsmath}
\usepackage{amssymb}
\usepackage{xspace}
\usepackage{bm}
\usepackage{color}
\usepackage[squaren,Gray]{SIunits}
\usepackage[hyperindex=true]{hyperref}
\hypersetup{linktocpage,colorlinks=true,citecolor=blue,linkcolor=blue}
\usepackage{empheq}
\usepackage{braket}
\usepackage{physics}

\makeatletter
\def\@bibdataout@aps{%
	\immediate\write\@bibdataout{%
		@CONTROL{%
			apsrev41Control%
			\longbibliography@sw{%
				,author="08",editor="1",pages="1",title="0",year="1"%
			}{%
				,author="08",editor="1",pages="1",title="",year="1"%
			}%
		}%
	}%
	\if@filesw \immediate \write \@auxout {\string \citation {apsrev41Control}}\fi 
}
\makeatother

\begin{document}

\title{Anyonic two-photon statistics with a semiconductor chip}	
	
\author{S. Francesconi}
\affiliation{Laboratoire Matériaux et Phénomènes Quantiques, Université de Paris, CNRS-UMR 7162, Paris 75013, France}

\author{A. Raymond}
\affiliation{Laboratoire Matériaux et Phénomènes Quantiques, Université de Paris, CNRS-UMR 7162, Paris 75013, France}
	
\author{N. Fabre}
\affiliation{Laboratoire Matériaux et Phénomènes Quantiques, Université de Paris, CNRS-UMR 7162, Paris 75013, France}
\affiliation{Centre for Quantum Optical Technologies, Centre of New Technologies, University of Warsaw, ul.  Banacha 2c, 02-097 Warszawa, Poland}
	
\author{A. Lemaître}
\affiliation{Université Paris-Saclay, CNRS, Centre de Nanosciences et de Nanotechnologies, 91120, Palaiseau, France}

\author{M. I. Amanti}
\affiliation{Laboratoire Matériaux et Phénomènes Quantiques, Université de Paris, CNRS-UMR 7162, Paris 75013, France}
	
\author{P. Milman}
\affiliation{Laboratoire Matériaux et Phénomènes Quantiques, Université de Paris, CNRS-UMR 7162, Paris 75013, France}
	
\author{F. Baboux}
\email{florent.baboux@u-paris.fr}
\affiliation{Laboratoire Matériaux et Phénomènes Quantiques, Université de Paris, CNRS-UMR 7162, Paris 75013, France}

\author{S. Ducci}
\affiliation{Laboratoire Matériaux et Phénomènes Quantiques, Université de Paris, CNRS-UMR 7162, Paris 75013, France}
	
\date{\today}

\begin{abstract}
	
Anyons, particles displaying a fractional exchange statistics intermediate between bosons and fermions, play a central role in the fractional quantum Hall effect and various spin lattice models, and have been proposed for topological quantum computing schemes due to their resilience to noise.
Here we use parametric down-conversion in an integrated semiconductor chip to generate biphoton states simulating anyonic particle statistics, in a reconfigurable manner. Our scheme exploits the frequency entanglement of the photon pairs, which is directly controlled through the spatial shaping of the pump beam.
These results, demonstrated at room temperature and telecom wavelength on a chip-integrated platform, pave the way to the practical implementation of quantum simulation tasks with tailored particle statistics.

\end{abstract}

\maketitle

Anyons, quasiparticles living in 2D or 1D with fractional statistics \cite{Wilczek82, Haldane91}, have stimulated strong research efforts over the past decades. This interest initiated with the theoretical prediction that anyons occur as elementary excitations in the fractional quantum Hall effect \cite{Halperin84,Arovas84}, in 2D electron gases submitted to a strong perpendicular magnetic field.
However, a direct observation of fractional statistics in these systems, in particular using single-particle interferometers \cite{Camino07,Willett13}, has long proven challenging, until the recent experimental demonstration of two-particle interference provides definitive evidence of anyonic statistics \cite{Bartolomei20}.

In parallel, anyons have been spotted as promising candidates to implement topological quantum computing and error correction tasks \cite{Nayak08}, due to their predicted robustness to noise. Among the variety of proposed schemes, models based on lattices of localized spins, as exemplified by the Kitaev toric code \cite{Kitaev03}, have attracted great attention. Indeed, it has been shown that the observation of anyonic excitations in the toric code does not necessitate a demanding ground-state cooling, but could also be obtained by generating dynamically the ground state and the excitations \cite{Han07}. Using this scheme, the fractional statistics of anyons and the topological path-independence of anyon braiding have been experimentally demonstrated using photons \cite{Pachos09,Lu09,Liu19}, cold atoms in optical lattices \cite{Dai17} and superconducting circuits \cite{Song18}.

The peculiar exchange statistics of anyons can also be simulated through the Hong-Ou-Mandel effect (two-particle interference) \cite{Hong87} of entangled photons. This has been demonstrated by implementing quantum walks in arrays of beamsplitters \cite{Sansoni12} or parallel waveguides \cite{Matthews13}, and by monitoring the propagation of spatial quantum correlations in scattering media \cite{vanExter12}. In all cases, these experiments rely on an external bulk source of photon pairs, which are then fed into a passive optical circuitry. By contrast, an integrated and reconfigurable source of entangled states with anyonic statistics would be a valuable asset in view of scalable applications in quantum information.

Here, we exploit spontaneous parametric down conversion (SPDC) in a semiconductor AlGaAs microcavity under transverse pumping \cite{Walton03,Caillet09,Orieux13} to engineer the exchange statistics of photon pairs directly at the generation stage, without post-manipulation. Our scheme exploits the frequency entanglement of the photon pairs, which is directly controlled through the spatial shaping of the pump beam. Applying a phase step profile to a Gaussian pump beam allows to tune the symmetry of the spectral wavefunction, so as to mimic any exchange statistics interpolating between bosonic and fermionic behaviors in two-photon interference; we exemplify this technique by simulating $\pi/2$ anyons. This constitutes to our knowledge the first demonstration of an integrated source of photon pairs with anyonic statistics.

\begin{figure*}[t]
\centering
\includegraphics[width=0.8\textwidth]{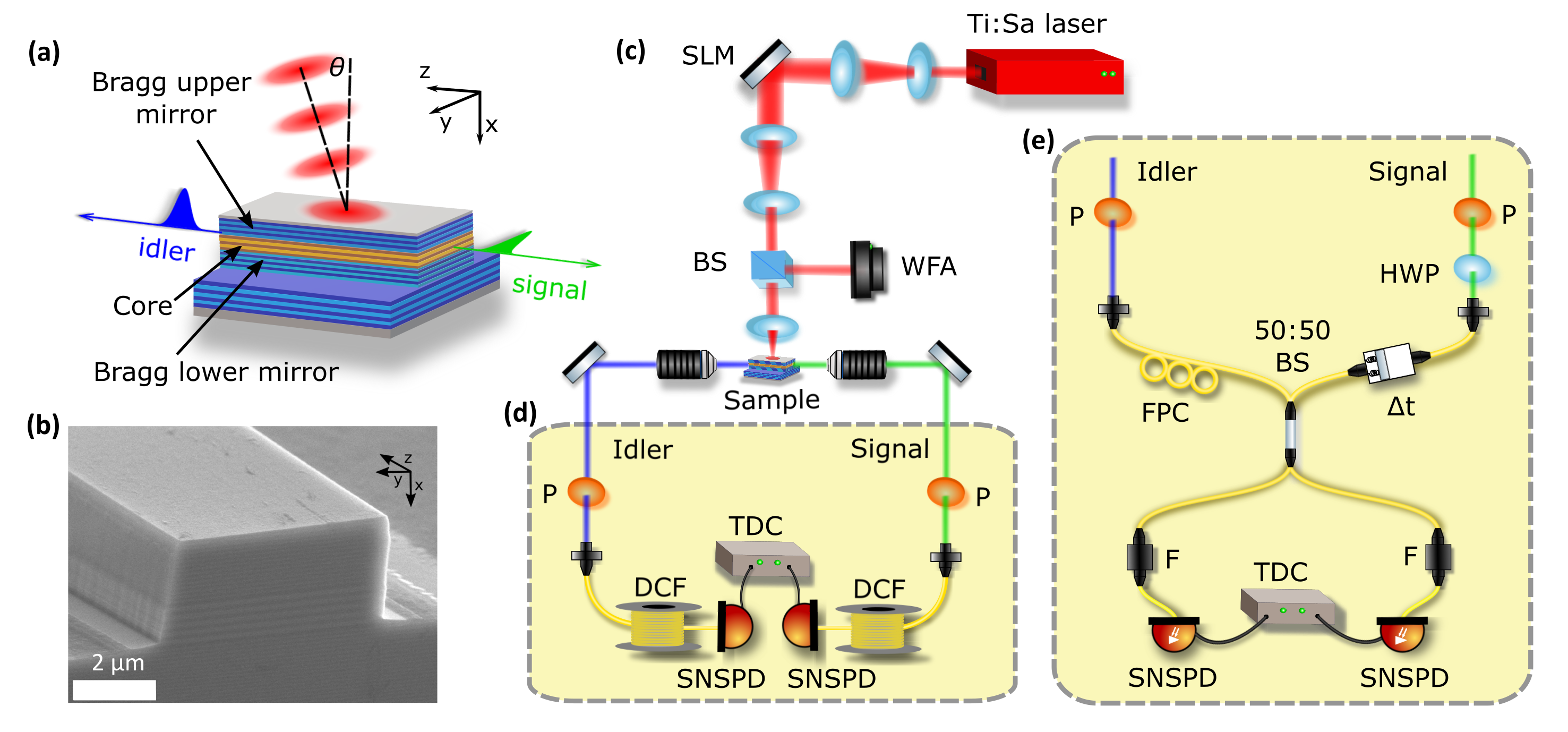}
\caption{
(a) Schematics and (b) SEM image of an AlGaAs ridge microcavity generating frequency-entangled photon pairs by SPDC in a transverse pump configuration.
(c)-(e) Sketch of the experiment, showing the pump shaping stage (c), the fiber spectrograph (d) and the Hong-Ou-Mandel (e) setups. Abbreviations: SLM=spatial light modulator, WFA=wavefront analyzer, BS=beam splitter, FPC=fibered polarization controller, P=polarizer, HWP=half-wave plate, F=frequency filter, DCF=dispersion compensating fiber, SNSPD=superconducting nanowire single photon detector, TDC=time-to-digital converter.
}
\label{Fig1}
\end{figure*}

\section*{Theory}

A schematics of our semiconductor integrated source of photons pairs is shown in Fig. \ref{Fig1}a, and a SEM image in Fig. \ref{Fig1}b. The source is a Bragg ridge microcavity made of a stack of AlGaAs layers with alternating aluminum concentrations \cite{Caillet09,Orieux11,Orieux13}. It is based on a transverse pump configuration, where a pulsed laser beam impinging on top of the waveguide (with incidence angle $\theta$) generates pairs of counterpropagating and orthogonally polarized photons (signal and idler) through SPDC \cite{DeRossi02,Orieux13}. 
Two nonlinear interactions can occur in the device \cite{Orieux13}; we consider here the one that generates a TE-polarized signal photon (propagating along $z>0$, see Fig. \ref{Fig1}a) and a TM-polarized idler photon (propagating along $z<0$).
The corresponding two-photon state reads
$\ket{\psi}= \iint d \omega_s d \omega_i  \phi(\omega_s,\omega_i) \hat{a}^\dagger_{s}(\omega_s) \hat{a}^\dagger_{i} (\omega_i)\ket{0,0}_{s,i}$,
where the operator $\hat{a}^\dagger_{s(i)}(\omega)$ creates a signal (idler) photon of frequency $\omega$.
The joint spectral amplitude (JSA) $\phi(\omega_s,\omega_i) $ is the probability amplitude to measure a signal photon at frequency $\omega_s$ and an idler photon at frequency $\omega_i$. Neglecting group velocity dispersion and considering a narrow pump bandwidth (which is justified in our experimental conditions), the JSA can be expressed as $\phi (\omega_s,\omega_i)=\phi_{\rm spectral}(\omega_+) \, \phi_{\rm PM}(\omega_-)$,
where we have introduced $\omega_\pm=\omega_s \pm \omega_i$ \cite{Boucher15,Barbieri17}.
The function $\phi_{\rm spectral}$, which reflects the condition of energy conservation, is given by the spectrum of the pump beam, while the function $\phi_{\rm PM}$, which reflects the phase-matching condition, is controlled by the spatial properties of the pump beam:
\begin{equation}\label{Eq2}
\phi_{\rm PM}(\omega_-)=\int_{-L/2}^{L/2} dz \,\mathcal{A}_p (z) e^{-i (k_{\rm deg}+\omega_-/v_{\rm g})z}
\end{equation}
Here, $\mathcal{A}_p (z)$ is the pump amplitude profile along the waveguide direction, $L$ is the waveguide length, $v_g$ is the harmonic mean of the group velocities of the SPDC photon modes and $k_{\rm deg}=\omega_p\text{\rm sin}(\theta_{\rm deg})/c$, with $\omega_p$ the pump central frequency, $c$ the velocity of light and $\theta_{\rm deg}$ the pump incidence angle that generates frequency-degenerate photons.
The direct relationship expressed by Eq. \eqref{Eq2} between the spatial profile ot the pump beam and the phase-matching of the SPDC process is a specific asset of the transverse pumping scheme, which offers a high versatility for the generation of frequency-entangled photonic states \cite{Caillet09,Boucher15,Francesconi20}.

Let us now show how we can exploit this versatility to simulate the exchange statistics of anyons.
We here concentrate on frequency-entangled photons, and show that they can behave in a HOM experiment exactly as anyons would, upon a proper engineering of the JSA (see Supporting Information).
To this aim, the JSA must be tailored such that:
\begin{equation}\label{eq:JSA}
	\phi(\omega_s,\omega_i)=e^{\pm i \alpha \pi} \phi(\omega_i,\omega_s) 
\end{equation}
The case $\alpha=0$ corresponds to bosons, $\alpha=1$ to fermions, while intermediate values of $\alpha$ define anyons. In the latter case, the wavefunction acquires a phase (different from $\pi$ or  $2\pi$) upon particle exchange, and the sign of this phase depends on the directionality of the exchange, which is a peculiarity of anyons. Consequently, exchanging anyons twice successively (an operation called braiding) does not lead back to the original wavefunction; rather, the wavefunction acquires a phase $e^{\pm i 2\alpha \pi}$ in this process. In a general manner, the number of exchange operations applied to a two-anyon wavefunction can be "counted" by the phase it acquires. Additionally, exchanging anyonic particles is equivalent to complex conjugation of the wavefunction \cite{Daletskii20,Santos12,Zaletel15}, meaning that in our case the JSA must verify $\phi(\omega_s,\omega_i)=\phi(\omega_i,\omega_s)^*$.

Since the spectral term $\phi_{\rm spectral}$ only depends on the frequency sum $\omega_+=\omega_s + \omega_i$, it is not affected by particle exchange, therefore Eq. \eqref{eq:JSA} reduces to a condition on the phase-matching function: $\phi_{\rm PM}(\omega_-)=e^{\pm i \alpha \pi} \phi_{\rm PM}(-\omega_-)$. We choose a suitable phase-matching function by analogy with the wavefunction describing the relative motion of two anyons placed in a harmonic potential \cite{Khare92}, since this situation yields Gaussian-like wavefunctions in close homology with the Gaussian frequency-time wavepackets naturally generated by SPDC. Translated into the frequency degree of freedom, the simplest choice of suitable phase-matching function is:

\begin{equation}
	\phi_{\rm PM}(\omega_-) \propto \omega_-^\alpha e^{-\omega^2_-/2 \beta^2}
\end{equation}
since $(-\omega_-)^\alpha=(e^{\pm i\pi})^\alpha (\omega_-)^\alpha $. Here, $\beta$ is a parameter controlling the width of the phase-matching condition. This choice for $\phi_{\rm PM}$ is well adapted to our experimental situation since, when pumping the ridge cavity with a Gaussian pump spot with flat phase profile, the resulting phase-matching function is purely Gaussian (i.e. $\alpha=0$) with a width $\beta=v_g / w_z$, where $w_z$ is the waist \cite{Orieux13,Boucher15}.

In the following, we focus on the simulation of anyonic particles with $\alpha=1/2$, i.e. acquiring a $\pi/2$ phase shift upon exchange, as occurs e.g. in the fractional quantum Hall effect at $1/2$ filling factor \cite{Halperin84,Arovas84}.
We target the implementation of the following phase-matching function:
\begin{equation} \label{eq:PhiPM}
	\phi_{\rm PM}(\omega_-) = \left\{
	\begin{array}{ll}
		\mathcal{C} e^{i\pi/4} \sqrt{\vert\omega_{-}\vert} e^{-\omega^2_-/2 \beta^2} & \text{if } \omega_{-} \geq 0\\
		\mathcal{C} e^{-i\pi/4} \sqrt{\vert\omega_{-}\vert} e^{-\omega^2_-/2 \beta^2} & \text{if } \omega_{-}<0
	\end{array}
	\right.
\end{equation}
where $\mathcal{C}$ is a real-valued normalization constant.
Note that this function verifies indeed $\phi_{\rm PM}(\omega_-)=e^{\pm i \pi/2} \phi_{\rm PM}(-\omega_-)$ (where the sign of the phase depends on the directionality of the exchange, which corresponds here to increasing or decreasing $\omega_-$) and  $\phi_{\rm PM}(\omega_-)=\phi_{\rm PM}(-\omega_-)^*$.

\begin{figure}[h]
\centering
\includegraphics[width=1\columnwidth]{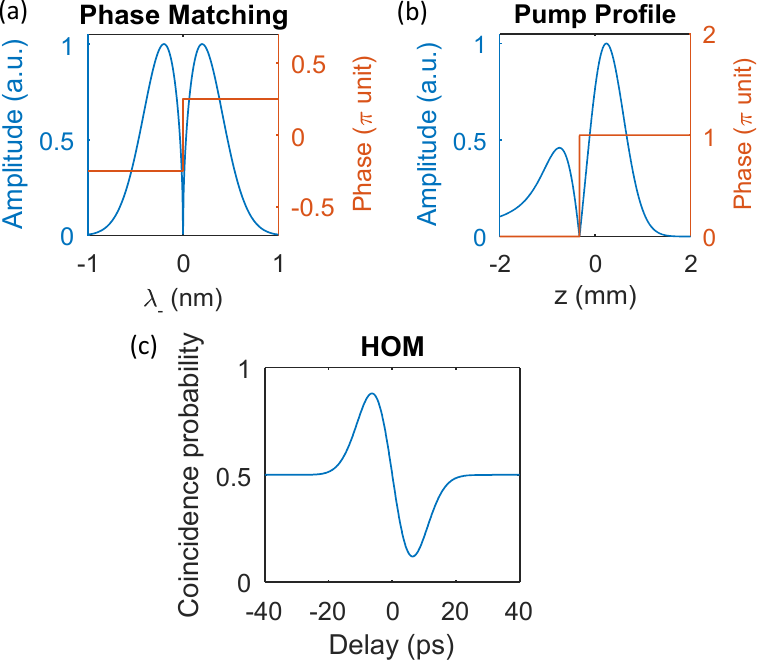}
\caption{
(a) Phase-matching function and (b) pump beam spatial profile (modulus in blue, phase in red) to simulate $\alpha=1/2 $ anyonic statistics with frequency-entangled biphotons. (c) Corresponding simulated Hong-Ou-Mandel interferogram.
}
\label{Fig2}
\end{figure}

Fig. \ref{Fig2}a shows a plot (in the wavelength space) of the phase-matching function of Eq. \eqref{eq:PhiPM} (modulus in blue, phase in red), with $\beta$ given by the parameters of our experiments presented in the following. This function consists of two symmetric lobes with a relative $\pi/2$ phase difference between them.
The pump beam spatial profile $\mathcal{A}_p(z)$ needed to generate such phase-matching function can be retrieved by inverse Fourier Transform of Eq. \eqref{Eq2}, $\mathcal{A}_p(z)\propto e^{i k_\text{deg} z} \int \dd \omega_- \: e^{i \omega_- z / v_g} \phi_{\rm PM }(\omega_-)$, which is valid if the pump beam size is narrower than the waveguide length $L$. The resulting pump spatial profile is plotted in Fig. \ref{Fig2}b (modulus in blue, phase in red): it consists of two uneven lobes, shifted from the waveguide center ($z=0$) and with a relative $\pi$ phase offset.

To reveal the symmetry of the entangled biphoton state encoded by the phase-matching function $\phi_{\rm PM }$, we resort to Hong-Ou-Mandel (HOM) interferometry. When two photons are delayed by a time $\tau$ and sent in the two input ports of a beamsplitter, they can either exit the beamsplitter through the same output port (bunching) or through opposite ports (antibunching), with two possibilities for each case. For independent and indistinguishable photons, antibunching probability amplitudes cancel each other at $\tau=0$; this leaves only bunching events, giving rise to the traditional HOM dip in the coincidence probability $P$ between the two beamsplitter outputs, typical of bosonic statistics ($\alpha=0$ in Eq. \eqref{eq:JSA}). For an entangled biphoton state, by contrast, the interference between the probability amplitudes associated to the four possible events can give rise to any value for the coincidence probability at zero delay, depending on the biphoton wavefunction \cite{Walborn03,Fedrizzi09,Sansoni10}. For instance, an entangled state antisymmetric with respect to particle exchange ($\alpha=1$) will give rise to a perfect coincidence peak ($P(\tau=0)=1$), as would be the case for independent fermions, while an entangled state with anyonic symmetry ($0<\alpha<1$) will produce a result intermediate between that of bosons and fermions.

Quantitatively, the HOM coincidence probability for frequency-entangled photons described by a phase-matching function $\phi_{\rm PM }$ can be calculated as:
\begin{equation} \label{EqHOM}
P(\tau) =  \frac{1}{2} \left( 1- \frac { {\rm Re} \left[\int \dd \omega_- \phi_{\rm PM }(\omega_-) \phi_{\rm PM }^*(-\omega_-) e^{-i \omega_- \tau}\right]}{\int \dd \omega_- \left\vert\phi_{\rm PM }(\omega_-)\right\vert^2 } \right)
\end{equation} 
Fig. \ref{Fig2}c reports the HOM interferogram calculated for the targeted anyons with $\alpha=1/2$. It shows a coincidence probability of $1/2$ at zero delay, a peak at negative delay and a dip at positive delay, and it is point-symmetric with respect to  $(\tau=0, P=1/2)$. 
These properties can be analytically retrieved by injecting the anyonic relationship $\phi_{\rm PM}(\omega_-)=e^{\pm i \alpha \pi} \phi_{\rm PM}(-\omega_-)$ into Eq. \eqref{EqHOM}, leading to  $P(\tau=0)= (1-\cos(\alpha \pi))/2$ and $P(\tau) -1/2 \propto \int_0^\infty \dd \omega_- \vert \phi_{\rm PM }(\omega_-) \vert^2 \cos(\omega_- \tau + \alpha \pi)$, which is an odd function of $\tau$ when $\alpha=1/2$.

The pump beam spatial profile (Fig. \ref{Fig2}b) needed to generate such biphoton state with anyonic statistics can be obtained using spatial light modulators (SLM). However, this requires a non-trivial control over both the intensity and phase of the beam. Here we will implement an easier situation, by using a single phase-only SLM. We use a pump beam having a Gaussian intensity profile shifted from the waveguide center, and with a $\pi$ phase step applied at $z=0$: numerical simulations (see Supporting Information) show that this pump spot produces a good approximation of the anyonic phase-matching function of Fig. \ref{Fig2}a, and thus a very similar HOM interferogram.

\section*{Experiments}

The experimental setup is sketched in Fig. \ref{Fig1}c. The AlGaAs device (waveguide length $L=1.9$ mm, width 6 $\mu$m and height 7 $\mu$m) is pumped with a pulsed Ti:Sa laser of central wavelength $\lambda_p \simeq 773 $ nm, pulse duration $\simeq$ 4.5 ps, repetition rate 76 MHz and average power $50$ mW on the sample. The pump beam is shaped with a reflective phase-only SLM in a 4f configuration, and analyzed with a Wavefront Analyser (WFA) to check the quality of the shaping. A cylindrical lens focuses the beam on the top of the waveguide, and the generated SPDC photons are collected with two microscope objectives and collimated into single-mode telecom optical fibers.

\begin{figure}[h]
	\centering
	\includegraphics[width=1\columnwidth]{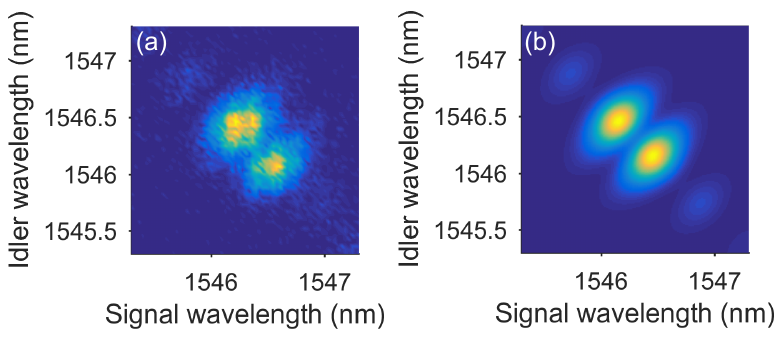}
	\caption{
		(a) Measured and (b) simulated joint spectral intensity of the generated biphoton state mimicking $\alpha=1/2$ anyons.
	}
	\label{Fig3}
\end{figure}

We set the pump incidence angle to $\theta_{\rm deg}\sim 0.5 \degree$ to produce frequency-degenerate photons \cite{Boucher15} and we pump the source with the aforementioned spatial profile, a Gaussian spot of waist $1$ mm, shifted from the waveguide center by $0.4$ mm and with a $\pi$ phase step at the center of the waveguide ($z=0$). We first characterize the emitted quantum state by measuring the Joint Spectral Intensity (JSI), that is the modulus squared of the JSA. For this we employ a fiber spectrograph \cite{Eckstein14} (see Fig. \ref{Fig1}d), where signal and idler are separately sent into a spool of highly dispersive fiber. This converts the frequency information into a time-of-arrival information, which is recorded using superconducting nanowire single photon detectors (SNSPD, of detection efficiency $80\%$) connected to a time-to-digital converter (TDC). 
Fig. \ref{Fig3}a reports the resulting measurement of the JSI. It shows two lobes of comparable intensity along the antidiagonal direction (corresponding to $\omega_-$), in accordance with the analytically calculated phase-matching function of anyonic-like biphotons (Fig. \ref{Fig2}a). The experimental JSI is also well reproduced by a full numerical simulation (Fig. \ref{Fig3}b) taking into account our experimental parameters (including the waveguide modal birefringence and chromatic dispersion) without adjustment, confirming the accurate tailoring of the joint spectrum through the pump spot profile. The generation rate of these quantum states is of $\simeq 10^7$ pairs/s at the chip output.

\begin{figure}[h]
	\centering
	\includegraphics[width=1\columnwidth]{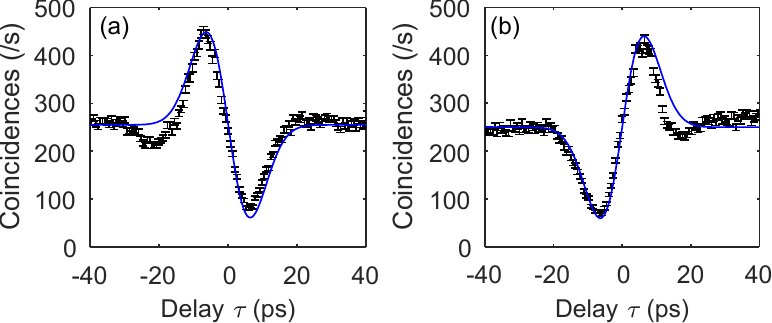}
	\caption{
		Measured (points) and simulated (lines) Hong-Ou-Mandel interferograms for a frequency-entangled biphoton state with $\alpha=1/2$ anyonic exchange statistics, for a pump spot with maximum intensity shifted at $z<0$ (a) and $z>0$ (b) along the waveguide direction. Experimental data show raw (uncorrected) counts.
	}
	\label{Fig4}
\end{figure}

We now perform two-photon interference in a HOM setup to reveal the underlying exchange statistics of the generated quantum state.
As sketched in Fig. \ref{Fig1}e, the polarization of the signal photon is rotated to align it with that of the idler, the signal photon then enters a fibered delay line and is recombined with the idler photon in a fibered 50/50 beamsplitter. Coincidence counts between the two output ports are recorded while varying the delay $\tau$ in the interferometer. 
The resulting HOM interferogram, shown in Fig. \ref{Fig4}a, displays a coincidence peak at negative delay and a dip at positive delay, in good quantitative agreement with the theoretical interferogram (solid lines) without adjustable parameter; the overlap between experiment and theory is $96.5$ \%.

Another implementation of $\alpha=1/2$ anyonic state can be obtained by using a pump spot profile symmetric to the previous one with respect to the waveguide center, i.e. with maximum intensity shifted by $+0.4$ mm instead of $-0.4$ mm. 
This corresponds to the other possible implementation of $\alpha=1/2$ anyons, with interchanged definitions for the sectors $\omega_{-}>0$ and $\omega_{-}<0$ in Eq. \eqref{eq:PhiPM}.
The corresponding HOM interferogram is shown in Fig. \ref{Fig4}b: it is mirror-symmetric to the interferogram of \ref{Fig4}a, with a coincidence dip at negative delay and a peak at positive delay. 
Again, good agreement is found with the theoretical inteferogram (solid lines) without adjustable parameter (overlap $97$ \%).

\section*{Discussion}

These results demonstrate the on-chip generation of biphoton states with anyonic exchange statistics, in a reconfigurable manner. This is a key asset of our approach compared to alternative recently reported schemes based e.g. on nonlinear domain engineering \cite{Graffitti18,Graffitti20}, that produce high-quality but non-reconfigurable frequency-entangled quantum states, or spectral shaping of the pump spot \cite{Mosley08,Kumar14,Ansari18}, which is reconfigurable but cannot modify the wavefunction symmetry and exchange statistics of the photon pairs. Also, in contrast to quantum walks of correlated photons \cite{Sansoni12,vanExter12,Matthews13} where the exchange statistics is studied through the spatial correlations at zero delay, here the temporal dependence of two-photon interference, with its peculiar point-symmetry typical of anyons, is revealed (Fig. \ref{Fig4}).

In conclusion, we have used an integrated semiconductor chip to simulate the particle statistics of anyons using frequency-entangled photon pairs generated by spontaneous parametric down-conversion, by engineering the spatial profile of the pump beam. Applying a phase step profile to a Gaussian pump beam allows to tune the symmetry of the spectral wavefunction so as to mimic the exchange statistics of anyons, as revealed by two-photon interference. This constitutes to our knowledge the first demonstration of an integrated and reconfigurable source of photon pairs simulating anyonic statistics. The photons are directly produced in two distinct spatial modes, at room temperature and telecom wavelength. The demonstrated source can be integrated monolithically with beamsplitters \cite{Belhassen18}, and the implementation of electro-optic phase shifters \cite{Wang14,Dietrich16} for state manipulation and superconducting nanowires for on-chip detection \cite{Schwartz18} can be envisaged. These assets open promising perspectives for the realization of quantum simulation tasks with tailored particle statistics in a chip-integrated platform \cite{Crespi13,Matthews13,Crespi15} without requiring external sources of quantum light, thus progressing towards real-world applications in quantum information.

\bigskip
\noindent\textbf{Acknowledgements}

We acknowledge support from European Union’s Horizon 2020 research and innovation programme under the Marie Skłodowska-Curie grant agreement No 665850, Paris Ile-de-France Région in the framework of DIM SIRTEQ (LION project), Ville de Paris Emergence program (LATTICE project), Labex SEAM (Science and Engineering for Advanced Materials and devices, ANR-10-LABX-0096), IdEx Université de Paris (ANR-18-IDEX-0001), and the French RENATECH network. N.Fabre acknowledges support from the project “Quantum Optical Technologies” carried out within the International Research Agendas programme of the Foundation for Polish Science co-financed by the European Union under the European Regional Development Fund.

\bibliography{D:/Travail/MPQ/Biblio}


\clearpage
\onecolumngrid
\begin{center}
	\textbf{\large Supporting Information}
\end{center}

\setcounter{equation}{0}
\setcounter{figure}{0}

\renewcommand{\theequation}{S\arabic{equation}}
\renewcommand{\thefigure}{S\arabic{figure}}

\subsection{Equivalence between HOM interference of anyons and entangled bosons}

This section demonstrates that the Hong-Ou-Mandel (HOM) interference of anyons can be simulated with entangled bosons, upon a proper engineering of their quantum correlations.

\bigskip

In the HOM experiment, the wavefunction of two particles of joint spectral amplitude $\phi(\omega_{s},\omega_{i})$ impinging on the beamsplitter (of respective spatial modes $a$ and $b$) with a delay $\tau$, can be written as:
\begin{equation}
	\ket{\psi(\tau)}=\iint d\omega_{s}d\omega_{i} \phi(\omega_{s},\omega_{i})e^{i\omega_{s}\tau} \hat{a}^{\dagger}(\omega_{s})\hat{b}^{\dagger}(\omega_{i})\ket{0,0}
\end{equation}
After the (balanced and lossless) beamsplitter and post-selecting only the coincidence events, we obtain:
\begin{equation}
	\ket{\psi(\tau)}=\frac{1}{2}\iint d\omega_{s}d\omega_{i} \phi(\omega_{s},\omega_{i})e^{i\omega_{s}\tau} ( \hat{a}^{\dagger}(\omega_{s})\hat{b}^{\dagger}(\omega_{i})-\hat{b}^{\dagger}(\omega_{s})\hat{a}^{\dagger}(\omega_{i}))\ket{0,0}
\end{equation}
The commutation relation of the operators depend on the particle statistics; they read, for bosons and anyons respectively:
\begin{align}
	\hat{b}^{\dagger}(\omega_{s})\hat{a}^{\dagger}(\omega_{i})=\hat{a}^{\dagger}(\omega_{i})\hat{b}^{\dagger}(\omega_{s})\\
	\hat{b}^{\dagger}(\omega_{s})\hat{a}^{\dagger}(\omega_{i})=A(\omega_{i},\omega_{s})\hat{a}^{\dagger}(\omega_{i})\hat{b}^{\dagger}(\omega_{s}) \label{comm_anyons}
\end{align}
where the function $A$ verifies $A(\omega_{i},\omega_{s})=A(\omega_{s},\omega_{i})^*$ and $\vert A \vert=1$; in the article we considered $A(\omega_{s},\omega_{i})=e^{i \alpha \frac{\pi}{2} \text{sign}(\omega_{s}-\omega_{i})}$.

The wavefunctions for the bosonic case ($\ket{\psi(\tau)}_{B}$, of joint spectrum $\phi_B$) and the anyonic case ($\ket{\psi(\tau)}_{A}$, of joint spectrum $\phi_A$) become, after performing a change of variable in the second term:
\begin{align}
	\ket{\psi(\tau)}_{B}=\frac{1}{2}\iint d\omega_{s}d\omega_{i} (\phi_B(\omega_{s},\omega_{i})e^{i\omega_{s}\tau}-\phi_B(\omega_{i},\omega_{s})e^{i\omega_{i}\tau}) \hat{a}^{\dagger}(\omega_{s})\hat{b}^{\dagger}(\omega_{i})\ket{0,0}\\
	\ket{\psi(\tau)}_{A}=\frac{1}{2}\iint d\omega_{s}d\omega_{i} (\phi_A(\omega_{s},\omega_{i})e^{i\omega_{s}\tau}-\phi_A(\omega_{i},\omega_{s})A(\omega_{i},\omega_{s})e^{i\omega_{i}\tau}) \hat{a}^{\dagger}(\omega_{s})\hat{b}^{\dagger}(\omega_{i})\ket{0,0}
\end{align}
The HOM coincidence probability $P(\tau)$ can then be calculated as $P(\tau)=\iint \abs{\bra{\omega_{s},\omega_{i}}\ket{\psi(\tau)}}^{2} d\omega_{s}d\omega_{i}$, yielding, for the bosonic and anyonic cases respectively:
\begin{align}
	P_{B}(\tau)=\frac{1}{2}\left(1-\text{Re}\left(\iint d\omega_{s}d\omega_{i} \phi_B(\omega_{s},\omega_{i})\phi_B^{*}(\omega_{i},\omega_{s})e^{i(\omega_{s}-\omega_{i})\tau}\right)\right)\label{coinboson} \\
	P_{A}(\tau)=\frac{1}{2}\left(1-\text{Re}\left(\iint d\omega_{s}d\omega_{i} \phi_A(\omega_{s},\omega_{i})\phi_A^{*}(\omega_{i},\omega_{s})A^{*}(\omega_{i},\omega_{s})e^{i(\omega_{s}-\omega_{i})\tau}\right)\right)\label{coinanyons}
\end{align}
where Re denotes the real part. 

Let us now assume that the joint spectral amplitude of the bosons $\phi_B$ is related to that of the anyons $\phi_A$ by  $\phi_B(\omega_s,\omega_i)=\sqrt{A(\omega_s,\omega_i)} \phi_A(\omega_s,\omega_i)$. Assuming that $\phi_A$ is symmetric under exchange ($\phi_A(\omega_i,\omega_s)=\phi_A(\omega_s,\omega_i)$), and given the properties of $A$ we find that $\phi_B$  verifies $\phi_B(\omega_i,\omega_s)=\phi_B^*(\omega_s,\omega_i)=A(\omega_i,\omega_s) \phi_B(\omega_s,\omega_i)$.
Using these properties, in Eq. \eqref{coinboson} we have $\phi_B(\omega_{s},\omega_{i})\phi_B^{*}(\omega_{i},\omega_{s})=\phi_B(\omega_{s},\omega_{i}) \phi_B^{*}(\omega_{s},\omega_{i})A(\omega_{s},\omega_{i})=\vert \phi_B(\omega_{s},\omega_{i}) \vert^2 A(\omega_{s},\omega_{i}) $
Similarly, in Eq. \eqref{coinanyons} we have $\phi_A(\omega_{s},\omega_{i})\phi_A^{*}(\omega_{i},\omega_{s})A^{*}(\omega_{i},\omega_{s})=\vert \phi_A(\omega_{s},\omega_{i}) \vert^2 A(\omega_{s},\omega_{i})$. Since $\vert A \vert=1$, $\vert \phi_B \vert=\vert \phi_A \vert$ and the two expressions of the coincidence probability are identical.

Therefore, the HOM interference of anyons with symmetric spectrum $\phi_A$ can be simulated using bosons with spectrum $\phi_B=\sqrt{A} \, \phi_A$. In the case of bosons, the term $\sqrt{A}$ in the spectrum mimicks the effect of fractional exchange statistics, which in the case of anyons is directly encoded in the operators algebra (Eq. \eqref{comm_anyons}).

\subsection{Calculation of HOM interferogram using the experimental pump spot profile}

\begin{figure*}[h]
	\centering
	\includegraphics[width=0.65\textwidth]{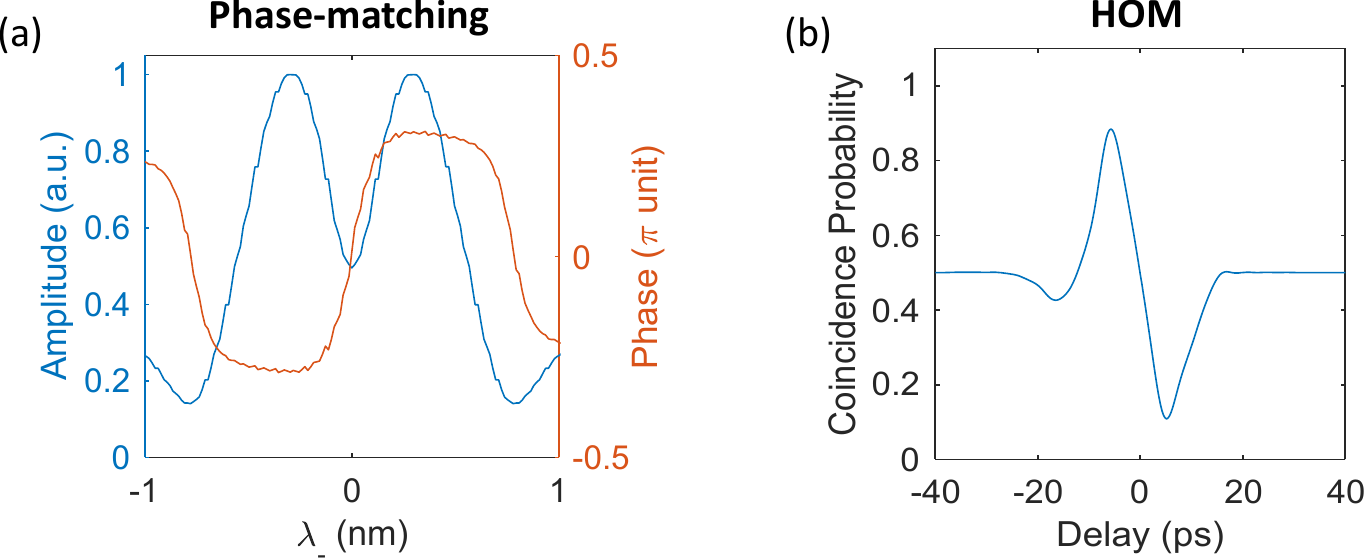}
	\caption{
		(a) Calculated phase-matching function and (b) corresponding HOM inteferogram for the pump spot profile experimentally implemented in the article to simulate $\alpha=1/2$ anyons.
	}
	\label{Fig_SM}
\end{figure*}

Figure \ref{Fig_SM} shows the phase-matching function (a) and corresponding HOM interferogram (b) calculated for the pump spot profile experimentally implemented in the article, i.e. a Gaussian spot of waist $1$ mm, shifted from the waveguide center by $0.4$ mm and with a $\pi$ phase step at the center of the waveguide. Similarly to the ideal case (Fig. \ref{Fig2}a of the article) the phase-matching function shows two symmetric lobes, with a close to $\pi/2$ phase step between them. The corresponding HOM interferogram is also close to the ideal one (Fig. \ref{Fig2}c of the article), albeit with a slight deviation near $\tau=-20 $ ps (consistently with the experimental data shown in Fig. \ref{Fig4} of the article).

\end{document}